\documentclass[useAMS,usenatbib,usegraphicx]{mn2e}

\usepackage{graphicx}
\DeclareGraphicsExtensions{.pdf,.png,.jpg,.ps}
\usepackage{epsfig}

\title[RR Lyraes in the LMC]{The Properties of the Large Magellanic Cloud Based on OGLE III Photometry of RR Lyrae Stars}
\author[R. Wagner-Kaiser, A. Sarajedini]
  {R.~Wagner-Kaiser,$^1$\thanks{Email: rawagnerkaiser@astro.ufl.edu}
  Ata~Sarajedini$^1$
\\
  $^1$University of Florida, Department of Astronomy, 211 Bryant Space Science Center, Gainesville, FL, 32611 USA\\
}

\begin{document}

\date{}

\pagerange{\pageref{firstpage}--\pageref{lastpage}} \pubyear{2002}

\maketitle

\label{firstpage}


\begin{abstract}
We present results from a study of ab-type RR Lyrae variables in the Large
Magellanic Cloud using the extensive dataset from phase III of the Optical 
Gravitational Lensing Experiment (OGLE). The metallicities of the RR Lyraes,
determined from the periods and amplitudes of their light curves, reveal a
statistically significant radial abundance gradient that is approximately one-half
of what is seen in the disks of the Milky Way and M33. The RR Lyrae abundance
gradient agrees with that of the old and metal-poor LMC globular clusters.
The reddenings of the OGLE RR Lyraes have been calculated using their
minimum light colors and reveal a mean value of E(V--I) = 0.12 $\pm$ 0.05,
where the quoted uncertainty represents the standard deviation of the mean.
The distribution of RR Lyrae extinctions across the face of the LMC is 
well-correlated with the distribution and emission intensity of CO clouds
based on recent millimeter wave surveys. In addition, we find that the old 
LMC globulars tend to be located in regions of low extinction. This underscores 
the need to survey the higher extinction regions with the specific aim of
increasing the sample of old LMC globular clusters. Finally, we examine the 
distance distribution of the RR Lyraes in order to probe the structure of the 
LMC and investigate the possibility that some of the RR Lyraes may reside in
a kinematically hot halo population. In addition to calculating a mean
LMC distance of $(m-M)_o$ = 18.55 $\pm$ 0.10 mag, we conclude that some
fraction of the RR Lyraes in our sample could be members of the LMC halo.
\end{abstract}



\section[]{Introduction}

The ancient stellar population of the Large Magellanic Cloud (LMC) is one
of the keys to unlocking the formation and early evolution of this very important
dwarf irregular galaxy. For decades, the metal-poor globular clusters in the
LMC have been used to study its early history because they represent distinct points in
space and time (van den Bergh 2004; Colucci et al. 2011). Knowledge of the ages, chemical abundances, and kinematical
properties of these clusters has led to a great many advances in our understanding
of the LMC's early evolution. However, based on the latest census, the LMC 
only contains 12 globular clusters (NGC 1466, 1754, 1786, 1898, 2210, 2257,
2005, 2019, 1835, 1841, Hodge 11, and Reticulum, excluding ESO121-SC09, 
Olsen et al. 1998; Johnson et al. 1999) with
ages comparable to those of the oldest Milky Way globulars. This relatively
small  number of objects limits their utility and applicability for the study of
galaxy-wide properties.

There is another tool at our disposal for studying the formation and early
evolution of the LMC - RR Lyrae variables (Smith 1995; Sarajedini 2011). 
These pulsating stars that are
located on the horizontal branch are ubiquitous in the LMC with tens of
thousands having been identified and characterized by recent surveys (Alcock
et al. 2000; Soszynski et al. 2009, see below).  
They are well known as distance indicators, and, in addition, their 
mere presence necessitates the existence of a stellar population
with an age greater than $\sim$10 Gyr.
Their periods and amplitudes are useful for measuring the metal
abundance of the star cluster or galaxy in which they reside.  Furthermore, their
minimum light colors can be used to measure the line-of-sight reddening.

There are three principal types of RR Lyrae variables; those pulsating in the
fundamental mode exhibit sawtooth-like light curves and are referred to as
ab-type or RR0 variables. The first overtone pulsators generally show sine-curve
shaped light curves, have shorter periods and typically lower amplitudes than the
ab-types, and are referred to as c-type or RR1 variables. Lastly, RR Lyraes that pulsate
in both the fundamental and first overtone modes (i.e. double mode pulsators) 
carry the d-type moniker (RRd).

The most recent survey of RR Lyraes in the LMC is represented by
the Optical Gravitational Lensing Experiment (OGLE, Udalski et al. 1992). We 
describe the detailed properties of this survey in the next section. For now, we
note that at least four studies have been published utilizing the vast
number of LMC RR Lyraes from OGLE. Pejcha \& Stanek (2009) used
9393 ab-type RR Lyraes in the OGLE-III catalog to investigate the
structure of the LMC stellar halo. After adopting a median distance of 50 kpc
for their sample of RR Lyraes, Pejcha \& Stanek (2009) construct a reddening
map of the LMC and use it to study the line of sight distribution of the
RR Lyraes. They find that the RR Lyraes define a triaxial ellipsoid with
axes ratios of 1:2.00:3.50 providing further evidence that the LMC does 
indeed possess a halo distribution of RR Lyrae variables (Kinman et al. 1991;
Minniti et al. 2003; Borissova et al. 2006). 

The series of papers by Haschke et al. (2011), Haschke et al. (2012a),
and Haschke et al. (2012b) used
the OGLE-III RR Lyrae data to explore the reddening, metallicity, and structure 
of the LMC, respectively. 
The first study noted above presented reddening maps of the LMC using the
red clump stars and the RR Lyrae variables and compared these maps
with several previous studies. The second paper, Haschke et al. (2012a),
used the Fourier properties
of the RR Lyrae light curves to study their metallicity distribution in the LMC
and compared the results with spectroscopy-based abundance values from
previous studies. The third paper, Haschke et al. (2012b), used the line-of-sight
distance distribution of the RR Lyraes and Cepheids to investigate the
structural properties of the LMC.
We will refer to these studies later as we compare our results with
theirs.

In the present work, we seek to study the reddening, metallicity, and
spatial distribution of the OGLE-III RR Lyraes using complementary techniques to
those employed in the previous studies of Pejcha \& Stanek (2009) and
Haschke et al. (2011, 2012a, 2012b). We will compare our results with 
those of previous
investigations in order to test the veracity of the techniques and the results they
yield. We also include the relevant properties of the old LMC globular clusters
in our analysis to gain some insight into their relationship with the RR Lyraes.
This paper is organized as follows.
The next section provides a brief description of the OGLE-III RR Lyrae
database. Section 3 presents our results on the metallicity, extinction, and
spatial distribution of the ab-type RR Lyraes and compares them
with previous studies. Lastly, Section 4 presents the conclusions
of this study.




\section[]{Data}

The OGLE experiment began its initial observations in 1992 with OGLE I. 
Since then, two additional datasets have been released (Udalski et al. 1992). 
The most recent one, from OGLE III, was taken from July 2001 to March 
2008 on the 1.3 meter Warsaw telescope at Las Campanas Observatory 
(Soszynski et al. 2009).  The imaging camera is composed of eight CCDs each with 
2048 by 4096 pixels providing a field of view of 35 by 35 arcmin on the sky 
(Udalski et al. 2003). Images were taken 
in the V and I filter passbands. The full photometric catalog including the
variable objects is available online. 

The reduction procedures, photometric calibrations, and astrometric 
transformations are detailed in Udalski et al. (2008). The raw light curves of
suspected variables were subjected to a period search algorithm based on 
Fourier analysis. From this, 24,906 RR Lyrae variables were identified and
characterized. For each object, the catalog provides  right
ascension, declination, mean magnitudes in I and V, period, I-band amplitude, 
along with the Fourier parameters $R_{21}$, $\phi_{21}$, $R_{31}$, and $\phi_{31}$
(Soszynski et al. 2009).

In the present work, we are mainly interested in the 17,693 ab-type RR Lyraes.
We have excluded those stars that were flagged as foreground objects and
those with uncertain or missing V-band photometry leaving 17,221 ab-type 
RR Lyraes, which constitute the dataset we will analyze herein.



\section{Results}

\subsection{Metallicities}\label{metallicities}

Metallicities for our sample are determined using the 
relationship from Alcock et al. (2000) shown in equation \ref{eq1}, wherein
P$_{ab}$ represents the period of the ab-type variables 
and V$_{amp}$ is the light curve amplitude in the V-band:

\begin{equation} \label{eq1}
[Fe/H] = -8.85 [log(P_{ab})+0.15 V_{amp}] - 2.6.
\end{equation}

\noindent To use equation \ref{eq1}, which is on the Zinn and West (1984) 
abundance scale, we need to convert the I-band amplitudes from the OGLE III
catalog to those in the V-band; for this, we make use of equation 1 in 
Dorfi \& Feuchtinger (1999). They find

\begin{equation} \label{eqn2}
V_{amp} = 0.075 + 1.497I_{amp},
\end{equation}

\noindent where V$_{amp}$ and I$_{amp}$ are the light curve amplitudes 
in the V- and I-band,
respectively. This empirical relation exhibits a R$^2$ correlation coefficient of 
0.904 according to Dorfi \& Feuchtinger (1999), where R$^2$ = 1 is a perfect
correlation. 

We note that Alcock et al. (2000) quote an error of 0.31 dex per star in their
metallicity-period-amplitude relation (equation 1). There is evidence that the 
actual error may be slightly less than this value and closer to 0.25 dex 
(Jeffery et al. 2011). There are at least two additional sources of error inherent
in equation (1). First, some RR Lyrae stars exhibit the Blazhko effect which results
in a modulation of their amplitude by as much as 0.2 mag (e.g. Kunder et al. 2010).
This leads directly to an uncertainty of 0.03 dex in their metal abundance. Second,
there is the conversion of the I-band amplitude to the V-band value. To estimate
the error inherent in this, we have examined the original data that were used to
construct equation (2) from Dorfi \& Feuchtinger (1999). We have calculated the 
root-mean-square (rms) deviation of the fitted points from the fit (i.e. equation 2)
and find a value of rms = 0.09 mag in the V band amplitude. Propagating this through 
the Alcock et al. equation yields a contribution of 0.014 dex to the total metal 
abundance error. Both of these sources of uncertainty are negligible and do not 
add significantly to the uncertainty of 0.31 dex originally quoted by Alcock et al. 
(2000).

\begin{figure}
  \centering
    \includegraphics[width=0.5\textwidth]{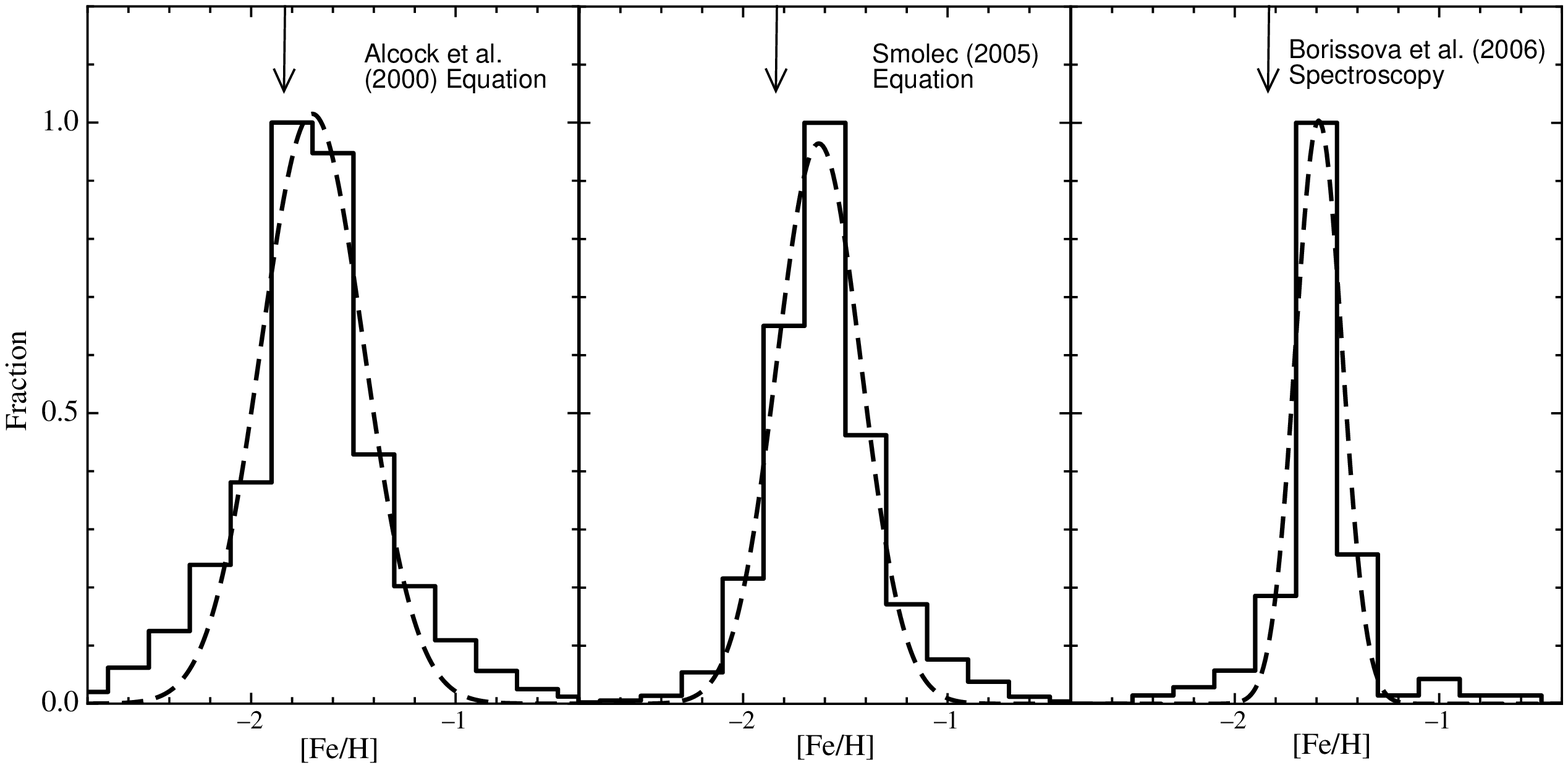}
  \caption{The left panel shows the histogram of metallicities from the present 
  work (solid line) along with a Gaussian fit to the distribution. These are based on
  the Alcock et al. (2000) equation. The middle panel shows the Fourier based 
  metallicities from Haschke et al. (2012a) using the Smolec (2005) relation.
  The right panel shows the spectroscopic values from Table 9 of 
  Borrisova et al. (2006). The vertical arrows represent the mean
  metal abundance of the 12 old globular clusters in the LMC. All of these values are on 
  the Zinn \& West (1984) scale.}
  \label{metallicityhist}
\end{figure}

The histogram of the resultant metallicities from the Alcock et al. (2000) relation
is shown in the left panel of Fig. \ref{metallicityhist} as the solid line. 
The middle panel of Figure \ref{metallicityhist} shows the metallicity
histogram derived using the methodology 
of Haschke et al. (2012a). Their study utilized the relationship derived by Smolec 
(2005) applied to the Fourier coefficients of OGLE RR Lyraes. The Smolec (2005)
relation is 

\begin{equation} \label{eq3}
[Fe/H]_{J95} = -3.142 -4.902 P_{ab} + 0.82  \phi_{31},
\end{equation} 

\noindent but it should be noted 
that a phase change must be accounted for in this equation, as the OGLE 
data uses a cosine Fourier series for their light curve fit while Smolec (2005) used 
a sine Fourier series.
Equation \ref{eq3} is on the Jurcsik (1995) metallicity scale. This was transformed to the 
Zinn \& West 1984 scale (ZW). 

\begin{equation} \label{eq4}
[Fe/H]_{ZW} = 1.028 [Fe/H]_{J95} - 0.242,
\end{equation}

\noindent which is taken from the paper by Papadakis et al (2000).
The histogram of metallicties resulting from equation \ref{eq4} is plotted in the 
middle panel of Figure \ref{metallicityhist} as the solid line.



The right panel of Fig. \ref{metallicityhist} illustrates the histogram of LMC RR Lyrae 
metallicities from Table 9 in the spectroscopic study of Borissova et al. (2006). 
These data were transformed from the Harris (1996) scale 
to the Zinn \& West scale (ZW) by adding --0.06 dex (Gratton et al. 2004;
Borissova et al. 2006).

\begin{table*}
\centering
    \begin{minipage}{180mm}
    \caption{Original LMC Globular Cluster Metallicity Data}
    \begin{tabular}{@{}cccccc@{}}
    \hline
Cluster & Suntzeff  & Grocholski & Mucciarelli & Colucci \\  
Name & et al. (1992) & et al. (2007) & et al. (2010) & et al. (2011) \\
\hline
  NGC 1466         & --1.85         & ...                         & ...        & ...           \\  
  NGC 1754         & --1.54         & ...                         & ...           & ...            \\  
  NGC 1786         & --1.87         & ...                         & --1.75        & ...            \\  
  NGC 1898         & --1.37         & ...                         & ...          & ...            \\  
  NGC 2210         & --1.97         & ...                         & --1.65        & ...            \\   
  NGC 2257         & --1.80          & --1.59                     & --1.95        & ...            \\  
  NGC 2005         & --1.92         & ...                         & ...           & --1.50         \\  
  NGC 2019         & --1.81         & --1.31                      & ...           & --1.61        \\  
  NGC 1835         & --1.79         & ...                         & ...        & ...            \\  
  NGC 1841         & --2.11         & ...                         & ...        & ...            \\  
        Hodge 11     & --2.06         & --1.84                    & ...           & ...            \\  
        Reticulum    & --1.71         & --1.57                    & É        & ...            \\ 
        \hline
    \end{tabular}
    \end{minipage}
\end{table*}

\begin{table*}
\centering
    \begin{minipage}{180mm}
    \caption{Adjusted LMC Globular Cluster Metallicity Data}
    \begin{tabular}{@{}ccccccl@{}}
    \hline
Cluster & Suntzeff & Grocholski & Mucciarelli  & Colucci & Average \\  
Name & et al. (1992) & et al. (2007) & et al. (2010) & et al. (2011) &  [Fe/H] \\
\hline
   NGC 1466         & --1.85         & ...          & ...           & ...            & --1.85                     \\ 
   NGC 1754         & --1.54         & ...          & ...           & ...            & --1.54                     \\ 
   NGC 1786         & --1.87         & ...          & --1.86        & ...            & --1.86   $\pm$ 0.01          \\  
   NGC 1898         & --1.37         & ...          & ...           & ...            & --1.37                     \\ 
   NGC 2210         & --1.97         & ...          & --1.78        & ...            & --1.87   $\pm$ 0.12          \\  
   NGC 2257         & --1.80         & --1.86       & --2.00        & ...            & --1.89   $\pm$ 0.07          \\  
   NGC 2005         & --1.92         & ...          & ...           & --1.81         & --1.87   $\pm$ 0.07          \\ 
   NGC 2019         & --1.81         & --1.60       & ...           & --1.92         & --1.78   $\pm$ 0.11          \\ 
   NGC 1835         & --1.79         & ...          & ...           & ...            & --1.79                     \\  
   NGC 1841         & --2.11         & ...          & ...           & ...            & --2.11                     \\  
       Hodge 11     & --2.06         & --2.07       & ...           & ...            & --2.06   $\pm$ 0.01          \\  
       Reticulum    & --1.71         & --1.85       & ...           & ...            & --1.79   $\pm$ 0.09          \\ 
        \hline
    \end{tabular}
    \end{minipage}
\end{table*}

Fitting a Gaussian function to each metallicity distribution in Figure \ref{metallicityhist}, 
we find the following 
peak values: --1.70 $\pm$ 0.25, --1.63 $\pm$ 0.21, --1.62 $\pm$ 0.10, 
for the present sample, the Haschke et al. (2012a) sample, and that 
of Borrisova et al. (2006),
respectively. The quoted uncertainties represent the 1-$\sigma$
standard deviations in each case. Given the errors, these values are in
statistical agreement, but it should be noted that the Fourier method of
Haschke et al. (2012a) yields metallicities with a slightly smaller
dispersion as compared with the Alcock et al. (2000) method that uses periods 
and amplitudes, which we have adopted. 

In order to compare the metallicities of the RR Lyrae variables with those
of the old globular clusters in the LMC, we have assembled abundance 
values from the literature for these 12 clusters as shown in Table 1. In order
to combine abundance values for the same clusters from different studies,
we need to ensure that they are on the same abundance scale. As such,
we have converted all of the metallicities to the Zinn \& West (1984) scale
and furthermore offset them to a common zero point. Since
Suntzeff et al. (1992) presents metallicities for all of the clusters, we have offset
the values from the other sources onto the system of Suntzeff et al. (1992)
and averaged the abundances for clusters in common between the various
studies. These adjusted values and the averages are shown in Table 2.
The errors in each average value have been calculated using the small
sample statistical formulae of Keeping (1962).

The vertical arrow in Fig. \ref{metallicityhist} shows the mean metal abundance 
of the 12 old globular clusters in the LMC taken from Table 2. 
This value equals [Fe/H] = --1.82 $\pm$ 0.06 (standard error of the mean, 
sem). We see that
this mean value is $\approx$0.3 dex more metal-poor than the field
RR Lyrae variables. However, given the uncertainties in both values, it
is unclear whether this difference of $\approx$0.3 dex is statistically
significant. If it is significant, then it could reflect a slight age difference
between the old globular clusters and the field RR Lyrae variables, in the
sense that the former would be older than the latter. 



\subsection{Metallicity Gradient in the LMC}\label{metallicitygrad}

\begin{figure}
  \centering
    \includegraphics[width=0.5\textwidth]{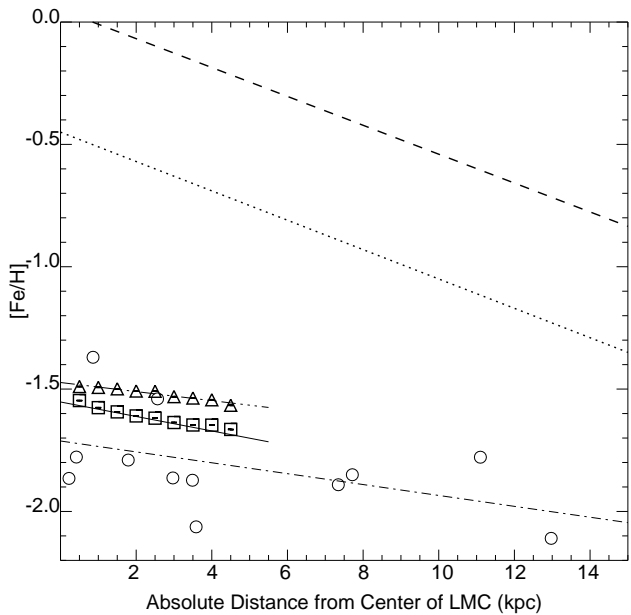}
  \caption{The radial metallicity gradient of the LMC is shown wherein the open 
  squares represent the binned radial distribution of the OGLE RR Lyraes in
  the present study while the open circles are the old LMC globular clusters from
  Table 2. The solid line is the least-squares fit to the RR Lyrae
  relation while the dot-dashed line is the fit to the globular clusters. The triangles
  represent the RR Lyrae abundance data from Hashcke et al. (2012a) binned in 
  an identical manner to our metallicity data with the dotted line representing a
  least squares fit to these points.  The metallicity gradients of M33 and the Milky
  Way are indicated by the dotted and dashed lines, respectively. The former is
  taken from the work of Tiede et al. (2004) while the latter represents the mean
  relation for the open clusters in the work of Friel et al. (2002).}
  \label{metallicitygradient}
\end{figure}

It is instructive to examine the radial metallicity variation among the stellar
populations of the LMC. For each RR Lyrae variable in our sample, we
calculate its angular distance from the center using equation (5), 

\begin{eqnarray} \label{angdist}
\cos{\theta}=[\sin{(90-\delta)}*\sin{(90-\delta_{LMC})}*\cos{(\alpha-\alpha_{LMC})}]  \\  \nonumber
+[\cos{(90-\delta)}*\cos{(90-\delta_{LMC})}] ,  \\  \nonumber
\end{eqnarray}

\noindent which is taken from van der Marel \& Cioni (2001). In this equation,
$\alpha_{LMC}$ and $\delta_{LMC}$ are the right ascension and declination of 
the LMC center from van der Marel \& Cioni (2001), while $\alpha$ and 
$\delta$ are the right ascension and declination of each OGLE RR Lyrae star.
A mean absolute LMC distance modulus of 18.55 (see Sec. 3.4) was used to 
calculate the radial location of each RR Lyrae from the center of the LMC. 
The metallicity data were then binned 
using 0.5 kpc bins and plotted in Fig. \ref{metallicitygradient}. 
The binned averages are plotted as squares along with a best fit line in
Figure \ref{metallicitygradient}.  This linear regression fit to the RR Lyrae 
data has a slope of --0.027 $\pm$ 0.002 dex/kpc, which is steeper than the
slope of --0.015 $\pm$ 0.003 dex/kpc found in Feast et al. (2010), who used the 
periods of the OGLE RR Lyraes along with
equation (1) of Sarajedini et al. (2006) in order to calculate metallicities. 
The open circles in Fig. \ref{metallicitygradient} are the old globular clusters in the
LMC as shown in the last column of Table 2. A least squares fit to these globular cluster
points yields a slope of --0.022 $\pm$ 0.013 dex/kpc. When the student's T-test is applied
to this latter value, we find a 93\% probability of significance suggesting that, given the
data in Table 2, the radial metallicity slope of the LMC globular clusters is not likely 
to be statistically significant.
We note that the open triangles in Fig. 2 represent the data from
Haschke et al. (2012a) binned in an identical manner to our data. The least
squares fit to these stars yields a slope of --0.019 $\pm$ 0.002 dex/kpc, which 
lies between the values from the present study and that of Feast et al. (2010).
It's not immediately clear why the techniques applied by Feast et al. (2010),
Haschke et al. (2012a), and the present work yield different radial metallicity gradients.
Further insights on which of these results is closest to reality will
require further spectroscopic metallicity measurements of individual RR Lyraes
in the LMC.

Also plotted in Fig. \ref{metallicitygradient} are the metallicity gradients in the 
disk of the Milky Way and M33 from Friel et al. (2002) and Tiede et al. (2004),
respectively. The former is based on the mean relation for all of the
open clusters in the study of Friel et al. (2002) while the latter is the mean
metallicity of RGB stars in an outer disk field of M33. These are meant
to illustrate a comparison between the disk metallicity gradients of these
three galaxies. In particular, it would appear that the LMC relation is significantly
shallower than those of the Milky Way and M33, which are both very similar. 

One reason for the different metallicity gradients in the disks of these
galaxies could be related to the ages of the stellar populations used to
measure these gradients. Friel et al. (2002) found that older open clusters
in the Milky Way's disk exhibit a significantly steeper radial metallicity
gradient as compared with younger clusters. This trend seems to be
reversed in the present case where the oldest stars in the LMC 
have a shallower metallicity gradient as compared with generally
younger populations in the Milky Way (open clusters) and M33 (RGB
stars). 

Another factor that could influence the slope of the radial metallicity gradient in
a galaxy is the presence or absence of a stellar bar. As Zaritsky et al. (1994, see also
Grocholski et al. 2007, Sharma et al. 2010) point out, barred spiral 
galaxies are more likely to possess shallower metallicity gradients as compared
with non-barred spirals. As such, this would explain the appearance of
Fig. 2, which suggests that the strong bar structure of the LMC has 
acted to diminish the extent of its disk metallicity gradient as compared with
M33, which has no measurable stellar bar, and the Milky Way, which possesses
a relatively small stellar bar.

%



\begin{table*}
\centering
    \begin{minipage}{180mm}
    \caption{LMC Globular Cluster Reddenings}
    \begin{tabular}{@{}cccl@{}}
    \hline
Cluster & E(B--V) & E(V--I) & Reference  \\  
\hline
  NGC 1466         &  0.09 &  0.12 & Johnson et al. (1999) \\  
  NGC 1754         &   0.09 & 0.12 & Olsen et al. (1998) \\  
  NGC 1786         &   0.07 & 0.09 &  Brocato et al. (1996)       \\  
  NGC 1898         &  0.07 & 0.09 & Olsen et al. (1998) \\  
  NGC 2210         &   0.06  & 0.08 & Brocato et al. (1996)   \\   
  NGC 2257         &  0.04 & 0.05 & Johnson et al. (1999)   \\  
  NGC 2005         &  0.10 &  0.13 & Olsen et al. (1998) \\  
  NGC 2019         &  0.06 & 0.08  & Olsen et al. (1998) \\  
  NGC 1835         &  0.08 &  0.10 & Olsen et al. (1998)\\  
  NGC 1841         &  0.18 &  0.23 & Brocato et al. (1996)   \\  
        Hodge 11     &   0.08 &  0.10 & Johnson et al. (1999) \\  
        Reticulum    &   0.05 & 0.07 & Mackey \& Gilmore (2004) \\ 
        \hline
    \end{tabular}
    \end{minipage}
\end{table*}


\subsection{Extinction Mapping}\label{extinction}

In order to determine the reddening for each RR Lyrae in our sample,
we make use of the results of Guldenschuh et al. (2005) who found
that the minimum light color of ab-type RR Lyrae variables is equal to 
$(V-I)_o$ = 0.58 $\pm$ 0.02 regardless of their intrinsic properties. 
As a result, we can calculate the reddening of an individual RR Lyrae 
via equation \ref{eq6} as

\begin{equation} \label{eq6}
E(V-I) = (V-I)_{min} - 0.58.
\end{equation}

\noindent To estimate the minimum light colors of the ab-type RR Lyraes in
the OGLE III database, we proceed as follows. If the light curves were symmetric
sine-like curves, we could calculate the minimum light color via the following equation:

\begin{equation} \label{eqn7}
(V-I)_{min} = [<V> + \frac{Amp(V)}{2}] - [<I> + \frac{Amp(I)}{2}].
\end{equation}

However, because the shapes of ab-type RR Lyrae light curves are 
asymmetric, we need to determine the correction to be applied
to the minimum light colors derived using equation (7). In order to calculate
this correction, we have used the ab-type RR Lyrae light curve templates
of Layden \& Sarajedini (2000) to calculate the quantity

\begin{equation} \label{eqn8}
\Delta = (V-I)_{min}^{template} - [<V> + \frac{Amp(V)}{2}] - [<I> + \frac{Amp(I)}{2}].
\end{equation}

\noindent The mean value of $\Delta$ for the 6 ab-type templates
is $\langle$$\Delta$$\rangle$ = -0.061 $\pm$ 0.017. As a result, we can correct
the minimum light colors derived from equation (7) by adding --0.061 mag to 
each of our values.



\begin{figure}
  \centering
    \includegraphics[width=0.5\textwidth]{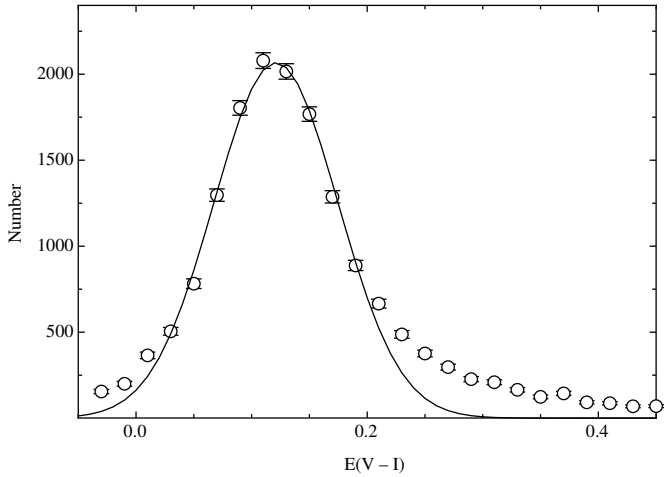}
  \caption{Histogram of  reddening values for LMC ab-type RR Lyraes in our data set. 
  The solid line is the Gaussian fit to the central portion of
  the distribution.}
  \label{redhist}
\end{figure}

This corrected apparent minimum light color is then coupled with the results of 
Guldenschuh et al. (2005) shown in equation (6) above to calculate the 
line-of-sight reddening. Figure ~\ref{redhist} illustrates the resultant 
reddening  distribution for the  LMC RR Lyraes in our sample. The number
of RR Lyraes exhibiting a negative reddening in Fig. ~\ref{redhist} 
represents only 5\% of the total, which is a testament to the robustness
of our method (Sarajedini et al. 2006). The error in each 
reddening is composed of $\sigma$=0.02 mag from the uncertainty in the value of 
$(V-I)_{o,min}$ and $\sigma$=0.017 mag from the correction applied to the
minimum light colors. Taken together, these combine to produce an uncertainty
of $\sigma$=0.026 mag in E(V--I). 

In order to test our method for reddening estimation, we have extracted the 
V and I band time-series photometric data for 200 RR Lyraes from the OGLE III survey. 
We have performed light curve fitting on these data (Layden \& Sarajedini 2000)
using the known periods in order to derive the minimum light colors of these stars. 
These were then used to calculate reddenings using equation 6 above. When we 
compare these reddenings to the results of applying our method (equations 7 and 8), 
we find a mean difference of 0.0291 $\pm$0.005 (sem) mag in E(V--I). This is
consistent with the error we have quoted for our reddening determinations
of $\pm$0.026 mag thus providing support for the technique we have used
to calculate the line of sight reddening to the LMC RR Lyraes.

A Gaussian profile was fit to the distribution of reddening values for the RR Lyrae. 
The peak of this distribution is at E(V--I) = 0.12 with $\sigma$ = 0.05, which
corresponds to E(B--V) = 0.09.  This value is in good accord with the work of 
Haschke et al. (2011), who found a mean LMC reddening of 
E(V -- I ) = 0.09 $\pm$ 0.07 mag from their analysis 
of red clump stars in the
LMC and E(V -- I ) = 0.11 $\pm$ 0.06 mag based on the OGLE RR Lyrae stars.
In fact, we can perform a star-by-star comparison of our reddenings with those
from Haschke et al. (2011). Doing this, we find a mean difference of 
-0.021 $\pm$ 0.041, which suggests that the two datasets are statistically
indistinguishable. This offset is also consistent with the quoted error of $\pm$0.026
in our values of E(V--I).
It should be noted that Haschke et al. (2011) utilized a completely different
method than we have to estimate the RR Lyrae reddenings. In particular,
they rely on RR Lyrae metallicities derived from the Fourier coefficients, which
are then used in conjunction with theoretical light curve models to infer
the intrinsic color of each RR Lyrae star in their sample. 

It is also interesting
to note that the mean reddening of the 12 old globular clusters in the LMC 
is E(V--I) = 0.11 $\pm$ 0.06 (see Table 3), which is consistent with that of the
RR Lyraes. All of these reddening values are, in turn, consistent with the range of
values published in Table 2 of Imara \& Blitz (2007), which extends from
E(V-I) $\approx$ 0.03 to E(V-I) $\approx$ 0.26 with a mean value of 
E(V-I) $\approx$ 0.16.

\begin{figure}
  \centering
    \includegraphics[width=0.5\textwidth]{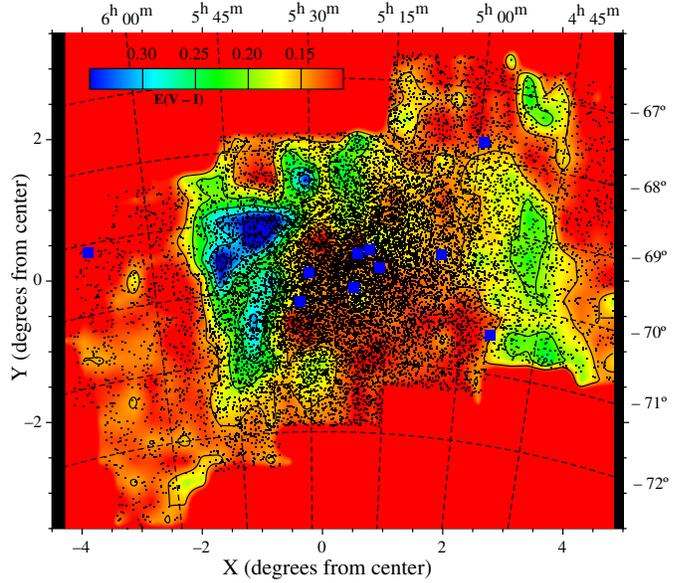}
  \caption{The contour map of extinction across the LMC based on the minimum
  light colors of the ab-type RR Lyraes from the OGLE III survey. The small points
  indicate the locations of the individual RR Lyrae variables while the blue squares
  are the old globular clusters in the LMC listed in Table 3 of Soszynski et al. (2009).}
  \label{RRLcontour}
\end{figure}

With reddenings for over 17,000 ab-type RR Lyraes in hand, we can also investigate
the extinction map of the LMC. This is shown in Fig. ~\ref{RRLcontour},
wherein the squares represent the locations of the old globular clusters
in the LMC from Table 1 of Soszynski et al. (2009). There are a number of
features of note in this diagram. First, the area around 30 Doradus and the
southern CO ridge are prominent sources of extinction in the case of the
RR Lyraes. There is an additional extinction ridge extending north-south along the
western edge of the LMC surveyed area at RA$\approx$4$^h$50$^m$.
It is also important to note that there is an extinction ``hole'' in the central
region of the LMC. Generally speaking, the oldest globular clusters in the LMC
(blue squares in Fig.  ~\ref{RRLcontour}) tend to lie in areas of reduced extinction. 
This phenomenon brings forth the question of whether undiscovered old clusters 
may exist in areas where high extinction has prevented their identification.
If so, it is vitally important for these old clusters to be identified and
characterized because
the overall dearth of old clusters in the LMC has made it 
challenging to study its early formation epochs using these clusters.

\begin{figure}
  \centering
    \includegraphics[width=0.5\textwidth]{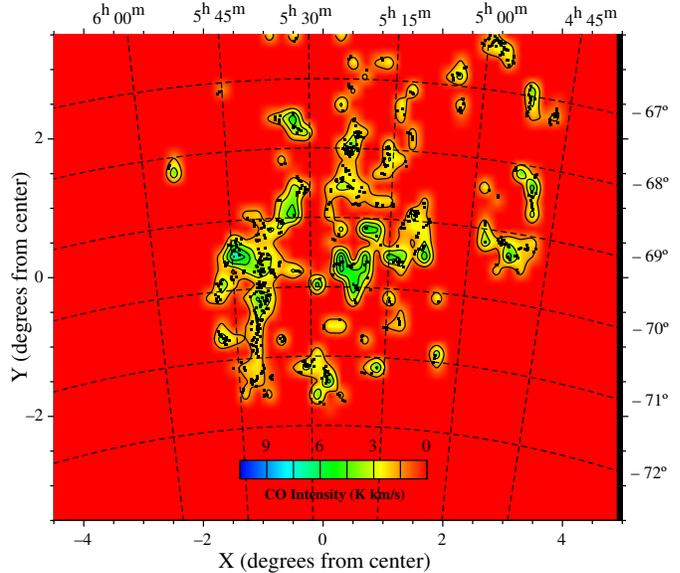}
  \caption{The contour map of LMC CO molecular cloud 
  intensites (K km/s) from the MAGMA
  survey of Wong et al. (2011). These are plotted on the same scale as the RR Lyrae
  reddening map.}
  \label{MAGMA2contour}
\end{figure}

\begin{figure}
  \centering
    \includegraphics[width=0.5\textwidth]{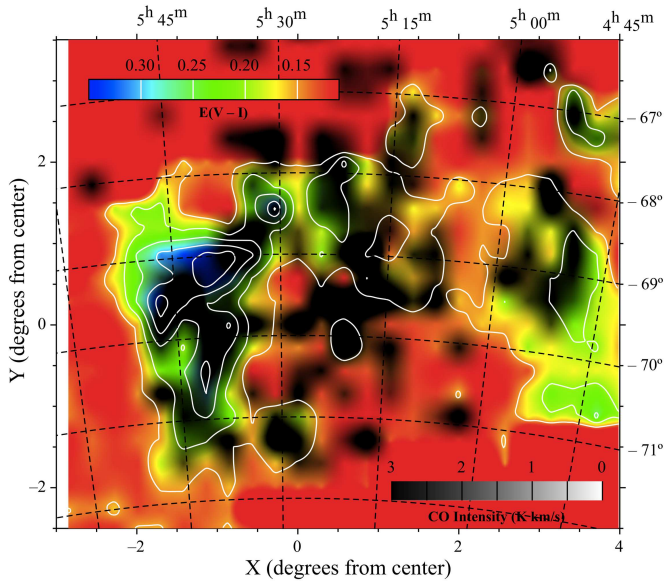}
  \caption{The RR Lyrae reddening map of the LMC is shown in color. Overplotted
  in greyscale is the contour map of CO molecular cloud 
  intensities (K km/s) from the work of Wong et al. (2011).}
  \label{RRL_MAGMAcontour}
\end{figure}

With a reddening map for the LMC in hand, it is of interest to compare it with
maps of the CO emission from the LMC. The latest version of such a map is
from the work of Wong et al. (2011) representing results of the Magellanic MOPRA 
Assessment (MAGMA) survey. This survey builds upon the previous work by 
Fukui et al. (2008) who used the 4.0m NANTEN radio telescope. 
Fig. ~\ref{MAGMA2contour}
shows the intensity in units of K km/s of the CO emission from the clouds
studied by Wong et al. (2011) from their Table 6. This figure is plotted to the same
scale as Fig.  ~\ref{RRLcontour}, which represents the RR Lyrae reddening
map.

A casual comparison of Fig. ~\ref{MAGMA2contour} and Fig. ~\ref{RRLcontour}
suggests a high degree of correlation between the extinction as revealed by
the RR Lyrae variables and the intensity of CO present in the line of sight.
Further insight into this correlation can be gained by overplotting the two 
distributions as shown in Fig. ~\ref{RRL_MAGMAcontour}. Once again, this
figure emphasizes that the RR Lyrae reddenings and the CO intensity distributions
are well correlated in their on-sky distribution. This suggests that the colors 
of the RR Lyraes are influenced by dust in the line of sight
further indicating that some of the RR Lyraes are located behind and within
the CO dust clouds. The one area that seems to be an exception to this
assertion is the central region of the LMC, where we see emission from CO 
clouds but where the RR Lyrae reddenings are all relatively small (see
Fig. ~\ref{RRL_MAGMAcontour}). It is possible that, in this area, the CO clouds
are predominantly on the far side of the LMC and/or the RR Lyraes are on
the near side so that the dust extinction does not influence the 
colors of the RR Lyrae stars.


\subsection{Distances}\label{distances}

With metallicities, reddenings, and mean magnitudes in hand, it is a simple matter to
calculate the distance to each of the over 17,000 ab-type RR Lyraes in the 
OGLE sample. First, we estimate the absolute magnitude of each RR Lyrae
using $M_V = (0.23\pm0.04)([Fe/H] + 1.6) + (0.56\pm0.12)$ from 
Chaboyer (1999), and calculate the
extinction using $A_V = 2.38 E(V-I)$ from Schlegel et al. (1998).





\begin{figure}
  \centering
    \includegraphics[width=0.5\textwidth]{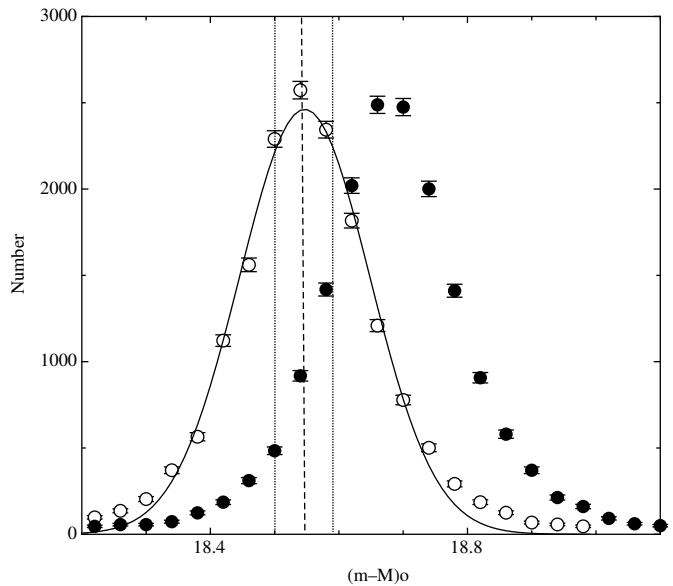}
  \caption{The open circles represent the histogram of distance moduli for the 
  RR Lyraes in our sample using the Chaboyer et al. (1999) relation between
  RR Lyrae absolute magnitude and metallicity. The solid line is the Gaussian 
  distribution fitted to the central portion of this histogram. The
  vertical dashed line represents the peak of the fitted Gaussian distribution
  while the dotted lines show the $\pm$1 kpc range of the putative LMC disk
  population (see text). In comparison, the filled circles represent the histogram of distance
  moduli yielded by the Benedict et al. (2011) relation between RR Lyrae absolute 
  magnitude and metallicity.}
  \label{disthist}
\end{figure}

The resulting distribution of RR Lyrae distance moduli is shown as the
open circles in Fig. ~\ref{disthist}.
A Gaussian profile fit to this distance distribution yields a peak
value of $(m-M)_o$ = 18.55 with a 1-$\sigma$ standard deviation of 0.10 mag. 
This translates to a distance of 51.3 $\pm$ 2.4 kpc. This value is well within
the range of values for the distance to the LMC presented in the
compilation of Clementini et al. (2003). Alternatively, we can perform this
calculation using the Benedict et al. (2011) calibration between RR Lyrae
absolute magnitude and metal abundance - 
$M_V = (0.214\pm0.047)([Fe/H]+1.5) + (0.45 \pm0.05)$. This
yields the binned histogram shown by the filled circles in Fig. ~\ref{disthist},
which has a peak value of $(m-M)_o$ = 18.68 with a 1-$\sigma$ standard 
deviation of 0.10 mag, as yielded by a Gaussian fit. Given the errors, this
distance modulus is consistent with the one derived from the Chaboyer (1999)
RR Lyrae relation, but it is on the high end of the distribution of LMC
distances from Clementini et al. (2003).

The question arises as to whether the line-of-sight dispersion in distance is
statistically significant or simply the result of measurement errors. To address
this question, we note that Alcock et al. (2000) estimate metallicity errors of
0.31 dex from equation (1) above. This translates to an error of $\sigma$=0.071 mag
in the absolute magnitude of each RR Lyrae. However, it is important to keep in
mind that the
1-$\sigma$ dispersion of the RR Lyrae metallicity distribution derived from the 
Alcock et al. (2000) relation (Fig. 1) is 0.25 dex, which is smaller than the error 
of an individual measurement of 0.31 dex. This suggests that 0.31 dex may be 
an overestimate of the error in equation (1). In any case, as discussed above, the 
reddening error is equal to $\pm$0.026 mag in E(V--I) leading to an uncertainty
of $\sigma$=0.062 mag in the determination of the
interstellar extinction. Ignoring for the moment the uncertainties in the measured 
V-band magnitudes, we find a distance modulus error of $\sigma$=0.094 mag.
This error is consistent with the dispersion measured in the distance histogram
shown in Fig. ~\ref{disthist} suggesting that this line of sight dispersion in distance
is dominated by the measurement errors in metallicity and reddening. 
We return to this point in the next section.




\subsection{Distance Along Line of Maximum Gradient}\label{maximumgradient}

To examine the structure of the LMC as revealed by the ab-type RR Lyraes, 
we can use the``kinematic circular disk method" as detailed in van der Marel \&
Cioni (2001) and revisited by Grocholski et al. (2007). In this method, one 
assumes a model of a circular disk with stars in circular and non-spherical orbits. 
Then, when plotting the line-of-sight distance versus the distance along the line 
of the maximum velocity gradient, objects following circular orbits should fall 
along a distinct line. The slope of this line should indicate the inclination of 
the disk. Thus, objects which deviate from this slope likely have non-circular orbits.

\begin{figure}
  \centering
    \includegraphics[width=0.5\textwidth]{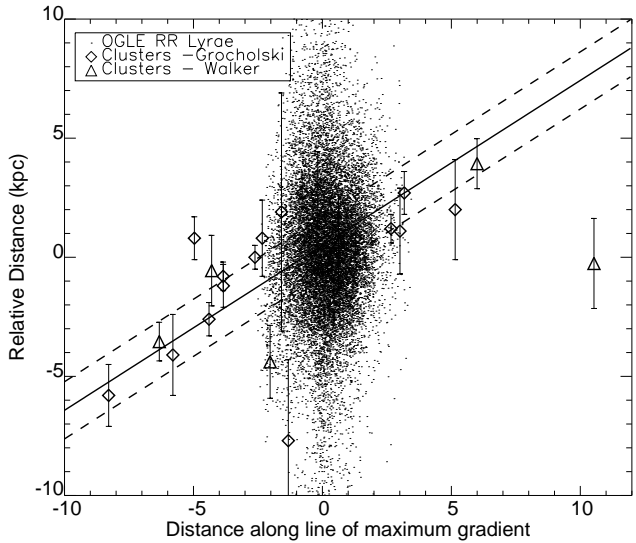}
  \caption{Line of sight distance versus distance along the line of maximum velocity gradient for a circular disk model (both in kpc). RR Lyraes from the present work are plotted as dots; old globular clusters are shown as diamonds (Grocholski et al. 2007) and triangles (Walker 1992). The solid line represents the expected slope of the LMC center. The dashed lines represents the 1 kpc thick disk of the LMC.}
  \label{maxgrad}
\end{figure}

In Figure ~\ref{maxgrad}, we show our derived distances for the RR Lyrae stars
as a function of the distance along the line of maximum velocity 
gradient. Also plotted are the clusters from Grocholski et al. (2007) and 
Walker (1992). Figure ~\ref{maxgrad} is modeled after Fig. 5 of Grocholski
et al. (2007), except that here we plot the line-of-sight distances relative to
the mean distance of the LMC. While the vast majority of the clusters in 
Figure ~\ref{maxgrad} are found to reside within 
$\approx$1 kpc of the LMC disk (as indicated by the dashed lines), the RR 
Lyrae variables do not appear to follow this trend. The distribution 
of RR Lyraes shows no indication of being oriented along the slope of the line 
(i.e. the inclination of the LMC disk). In addition, a significant number of the
RR Lyraes ($\approx$45\%) are found in the region beyond the 1 kpc thick disk 
(see Fig. ~\ref{disthist}) assumed for the LMC (Grocholski et al. 2007). 
One interpretation of this result is that the RR Lyraes in the OGLE
sample populate a kinematically hot halo around the LMC, consistent with the 
findings of Minniti et al. (2003) and Pejcha \& Stanek (2009, for a
contrasting view, see Subramaniam \& Subramanian 2009).
However, as discussed in the previous section, the line-of-sight distance 
distribution of the RR Lyrae seems to be dominated by the measurement errors. 
As a result, the value of $\approx$45\% is likely to be an
upper limit for the fraction of the OGLE ab-type RR Lyraes that are
members of the LMC halo.

\section{Conclusions}\label{conclusions}

In this paper, we present metallicities, reddenings, and distances of RR Lyrae stars
in the LMC based on the OGLE III database. These stars are used to 
examine the metallicity gradient of the LMC, its extinction profile, and 
the line-of-sight distribution of the RR Lyrae as a proxy for the LMC's
halo population. From this investigation, we draw the following conclusions:

1. There appears to be a radial metallicity gradient of the RR Lyraes in the LMC. 
The significance of the gradient depends on how the properties of the RR Lyrae 
light curves (e.g. periods, amplitudes, Fourier coefficients) are converted to 
metallicity. Given this apparent dependence, it is important to
measure spectroscopic metal abundances for a large sample of LMC RR Lyraes
with a substantial radial extent.

2. The extinction of the RR Lyraes in the LMC is correlated with the
location and emission of the CO clouds detailed by previous studies. 
However, the CO clouds do not appear to be correlated with the density of 
RR Lyraes across the face of the LMC. In addition, the old LMC globulars
tend to be located in regions of low extinction. This underscores the need
to survey the regions of higher extinction searching specifically for more old
globular clusters.

3. Our analysis of the OGLE III ab-type RR Lyraes is inconclusive with
regard to whether these stars belong to the halo or disk of the LMC.
The RR Lyraes appear to have a line-of-sight distribution that is not
consistent with the expected behavior of the LMC disk. However, the
uncertainties inherent in our analysis and the distribution of the OGLE RR Lyraes
in the LMC render this a tentative conclusion. In any case, previous work, such
as Minniti et al. (2003), does suggest that there are RR Lyraes in the LMC
that belong to a kinematically hot halo.


\section*{Acknowledgments}

We thank Tony Wong for assistance with the MAGMA CO maps. We are also
grateful to Aaron Grocholski for help with various aspects of the analysis
presented in this paper. Both Raoul Haschke and Karen Kinemuchi provided
valuables comments on an early version of this manuscript that greatly improved
its clarity and presentation. We also thank an anonymous referee whose comments
and suggestions were very helpful.

\bsp

\label{lastpage}

\end{document}